# Edge Triggering in IoT Mesh Networks: A Comparative Monte Carlo Study of Seven Detection Algorithms

*Sergii Makovetskyi*, *Lars Thomsen*[*]

*Abstract:* Real-time event detection in Internet of Things (IoT) mesh sensor networks presents significant challenges due to time-varying noise conditions, limited computational resources at edge nodes, and the need for autonomous operation without centralised coordination. This paper presents a comprehensive Monte Carlo simulation study comparing the Temporal Spectral Noise-Floor Adaptation (TSNFA) method against six alternative detection algorithms, evaluated across a 200-node mesh network over 24 hours with realistic noise models including 60 Hz electromagnetic interference (EMI), sinusoidally drifting noise power (±6 dB), and intermittent digital switching bursts. TSNFA achieves 100% detection rate with zero false positives, uniquely combining three interlocking defences: spectral band selection, temporal persistence filtering, and adaptive noise-floor tracking. Every competing algorithm omits at least one of these three defences and fails correspondingly, with false-positive rates ranging from 0 (Send-on-Delta, which also detects nothing) to 13,387,930 (broadband energy ratio). These results identify the three-defence combination as necessary and sufficient for autonomous edge triggering in resource-constrained IoT deployments.

## I. INTRODUCTION

The proliferation of Internet of Things (IoT) sensor networks has enabled unprecedented monitoring capabilities across diverse domains including environmental sensing, structural health monitoring, industrial process control, and precision agriculture [1]. These networks typically comprise numerous low-power sensor nodes communicating through mesh topologies, where each node must autonomously detect events of interest and transmit relevant data to a central sink node for aggregation and analysis [2], [3].

A fundamental challenge in such deployments is the reliable detection of genuine events amidst time-varying background noise. Real-world sensor environments exhibit complex noise characteristics including electromagnetic interference (EMI) from power systems, environmental disturbances from wind and precipitation, and mechanical noise from motors and machinery. The noise floor at any given node fluctuates significantly over time, rendering fixed-threshold detection approaches ineffective.

The consequences of suboptimal event detection are severe in both directions. False negatives which are missed genuine events compromise the monitoring system's primary function. False positives which are spurious triggers on noise consume limited network bandwidth, drain battery resources through unnecessary transmissions, and can overwhelm downstream processing systems and causes distrust in the results from the system. In mesh networks where nodes relay messages for their neighbours, false-positive cascades can cause network congestion and collapse.

Existing approaches to adaptive thresholding predominantly operate in the time domain, adjusting detection thresholds based on running estimates of signal amplitude statistics. While computationally simple, these methods cannot distinguish between noise energy and signal energy when both occupy similar amplitude ranges. The fundamental limitation is that time-domain analysis conflates all energy sources, making it impossible to selectively respond to event signatures while rejecting noise.

In our previous work [4], we introduced Temporal Spectral Noise-Floor Adaptation (TSNFA), a frequency-domain approach that exploits the distinct spectral characteristics of genuine events versus noise sources. By performing lightweight spectral analysis and maintaining adaptive thresholds in specific frequency bands, TSNFA can detect low-amplitude events that would be masked by noise in time-domain analysis, while rejecting high-amplitude noise that falls outside the event frequency band.

This paper presents a rigorous Monte Carlo simulation study comparing TSNFA against six competing detection algorithms spanning time-domain methods, fixed-threshold spectral methods, broadband energy detectors, protocol-level data-reduction schemes, and a TinyML autoencoder.

Our contributions are:

1. A comprehensive simulation framework incorporating Poisson-distributed events, multi-component noise models, and 200-node mesh network over a 24-hour duration;

2. Quantitative performance comparison demonstrating TSNFA's 100% detection rate and zero false positives versus false-positive rates of up to 13.4 M for the worst competing method;

3. A decomposition showing three specific signal processing defenses:
   - spectral band selection,
   - persistence filtering,
   - adaptive noise-floor tracking

are together necessary and individually insufficient for reliable detection in this environment.

[*] *S. Makovetskii is a Ph.D. student at Kharkiv National University of Radio Electronics, Ukraine (e-mail: serhii.makovetskyi@nure.ua). L. Thomsen is Managing Director of Gnacode Inc., Medicine Hat, AB, Canada (e-mail: lt@gnacode.com).* *This work was supported in part by Gnacode Inc. under grant 1037487 by the National Research Council of Canada.*

## II. RELATED WORK

### *A. Event Detection in Wireless Sensor Networks*

Event detection in wireless sensor networks has been extensively studied, with approaches broadly categorised into centralised and distributed methods [2]. Centralised approaches collect raw data at a fusion centre for joint processing, enabling sophisticated detection algorithms but incurring high communication costs and latency. Distributed approaches perform detection at individual nodes, transmitting only when events are detected, but face challenges in maintaining detection quality with limited local information [5].

### *B. Adaptive Thresholding Methods*

Recognising the limitations of fixed thresholds, researchers have developed various adaptive approaches. Moving-average methods adjust thresholds based on recent signal history. Exponentially-weighted moving-average (EWMA) schemes provide smoother adaptation with tunable responsiveness. Zhang et al. [6] proposed an influential adaptive-threshold method using exponential smoothing with a trigger multiplier of 3× the adaptive threshold. While achieving significant bandwidth reduction, the method inherently cannot distinguish signal from noise based on spectral content. Hussein et al. [7] proposed the Dual-Energy Dynamic-Range (DEDaR) detector, which triggers on broadband energy ratio. Correa et al. [8] applied Send-on-Delta (SoD) for IoT data-reduction. Each of these methods is evaluated in the comparative study in Section III.

*C. Frequency-Domain Approaches* Spectral analysis for event detection has received less attention in the IoT literature, primarily due to computational constraints at resource-limited nodes. Recent advances in efficient spectral estimation have renewed interest in frequency-domain methods. The Goertzel algorithm enables computation of individual frequency bins without full FFT overhead [9]. Bhoi et al. [10] applied STFT-based spectral gating to IoT sensor data. Our TSNFA method builds on these advances, implementing lightweight spectral analysis optimised for edge deployment while maintaining adaptive capabilities for time-varying noise environments.

*D. TinyML and Neural Approaches* Hammad et al. [11] proposed TinyML autoencoder anomaly detection for resource-constrained nodes. The approach trains a neural network to learn normal noise patterns and flags reconstruction errors above a learned threshold as anomalies. While attractive for complex pattern spaces, the method shares the fundamental weakness of all fixed-reference detectors: it cannot adapt its threshold to a slowly drifting noise environment without retraining, which is prohibitively expensive on Cortex-M0+ class hardware.

## III. SIMULATION FRAMEWORK AND RESULTS

This section formalises the detection algorithms evaluated in the comparative Monte Carlo study. For each method we first provide a plain-language overview and the governing equations, then the pseudocode implementation, and finally a line-by-line annotation that traces the signal-processing rationale behind every operation. The goal is to make the theoretical basis fully reproducible.

All algorithms operate on a discrete-time signal x[n] sampled at $f_s = 100$ Hz and segmented into non-overlapping frames of $L = 128$ samples, giving a frame duration $T_\beta = L/f_s = 1.28$ s and a frame rate $f^{fr}{}_{ame} \approx 0.781$ Hz. The frequency resolution of the L-point FFT is $\Delta f = f_s/L = 0.781$ Hz, placing the selected event band [1, 5] Hz in FFT bins K = {1, 2, 3, 4, 5, 6}.

*Monte Carlo signal model.* The simulation evaluates all algorithms against a common synthetic signal created to represent a realistic signal experienced by the sensors when deployed outdoor. The sampled signal at node i and sample index n within frame m is

$$x_i[n] = s[n - \tau_i] + w_{th}[n] + w_{EMI}[n] + w_{dig}[n] \quad (1)$$

where s[n] is the event waveform (a damped sinusoidal impulse at 1–5 Hz with SNR = 18 dB relative to in-band noise power P), $w_{th}$ is zero-mean white Gaussian thermal noise with power spectral density $P/f_s$, $w_{EMI}$ is 60 Hz mains interference at amplitude $0.3\sqrt{P}$, modelled as a simplified fixed fraction of the noise amplitude to represent worst-case EMI coupling that co-varies with the receiver noise floor, and $w_{dig}$ equally simplified represents intermittent digital switching bursts at 800–2000 Hz with amplitude up to $2.0\sqrt{P}$. The noise power P itself varies sinusoidally over a 1-hour period with ±6 dB excursion:

$$P(t) = P_0 \cdot 10^{(A/10)\,\sin(2\pi t\,/\,T_{\text{cycle}})} \quad (2)$$

where A = 6 dB and $T_{cycle} = 3600$ s, so that P swings between $P_0/4$ and $4P_0$ (a factor of 16 from minimum to maximum) over each hour. The chosen noise model introduces three failure modes that challenge the different algorithms tested in the simulation: the ±6 dB drift forces any viable detector to adapt its threshold over time, the 60 Hz EMI injects structured interference that a time-domain detector cannot distinguish from event content, and the intermittent digital bursts create impulsive transients at unpredictable intervals. Each algorithm receives the identical signal realisation; performance differences arise solely from the detection logic. **Figure 1** illustrates the data flow of the Monte Carlo simulation framework used to evaluate Algorithm 1 (TSNFA-mean, Makovetskii & Thomsen 2026) against Algorithm 3 (Zhang et al. 2023). True events are scheduled via a Poisson process and stored in true_event_times[node_id]. A NoiseGenerator synthesises realistic interference as: 60 Hz EMI harmonics, wind, rain, and motor noise. All which are combined with the event signal inside _generate_frame_samples() to produce a 128-sample frame of noise plus event signal. This frame is injected in parallel into both detection algorithms. Algorithm 1 computes the FFT magnitude spectrum, applies a temporal mean filter over γ_d spectra to suppress transients, then compares the in-band energy against an adaptive noise-floor threshold ζ·N_k across the 1–5 Hz band. Algorithm 3 instead applies a sample-by-sample EMA threshold directly in the time domain with no frequency filtering. Both algorithms report a trigger time, which _is_true_positive() cross-references against the scheduled event window to classify each trigger as a True positive or False Positive, producing the detection rate and false alarm rate reported in the results.

### *A. Algorithm 1 TSNFA-mean: Temporal Spectral Noise-Floor Adaptive Triggering (Mean Variant)*

*1) Overview*

The Temporal Spectral Noise-Floor Adaptive (TSNFA) trigger is the detection method proposed by Makovetskii and Thomsen 2026. It is built on three interlocking defences, each targeting a distinct failure mode observed in deployed IoT perimeter-security sensors.

*Defence 1: Spectral band selection.* An L-point FFT decomposes each frame into frequency bins, and only the bins covering the event band $[f_{low}, f_{high}] = [1, 5]$ Hz are retained. All energy from noise sources like electromagnetic interference (EMI) at 60 Hz and its harmonics, digital switching noise above 100 Hz, and broadband thermal noise outside the event band is discarded before any detection statistic is computed.

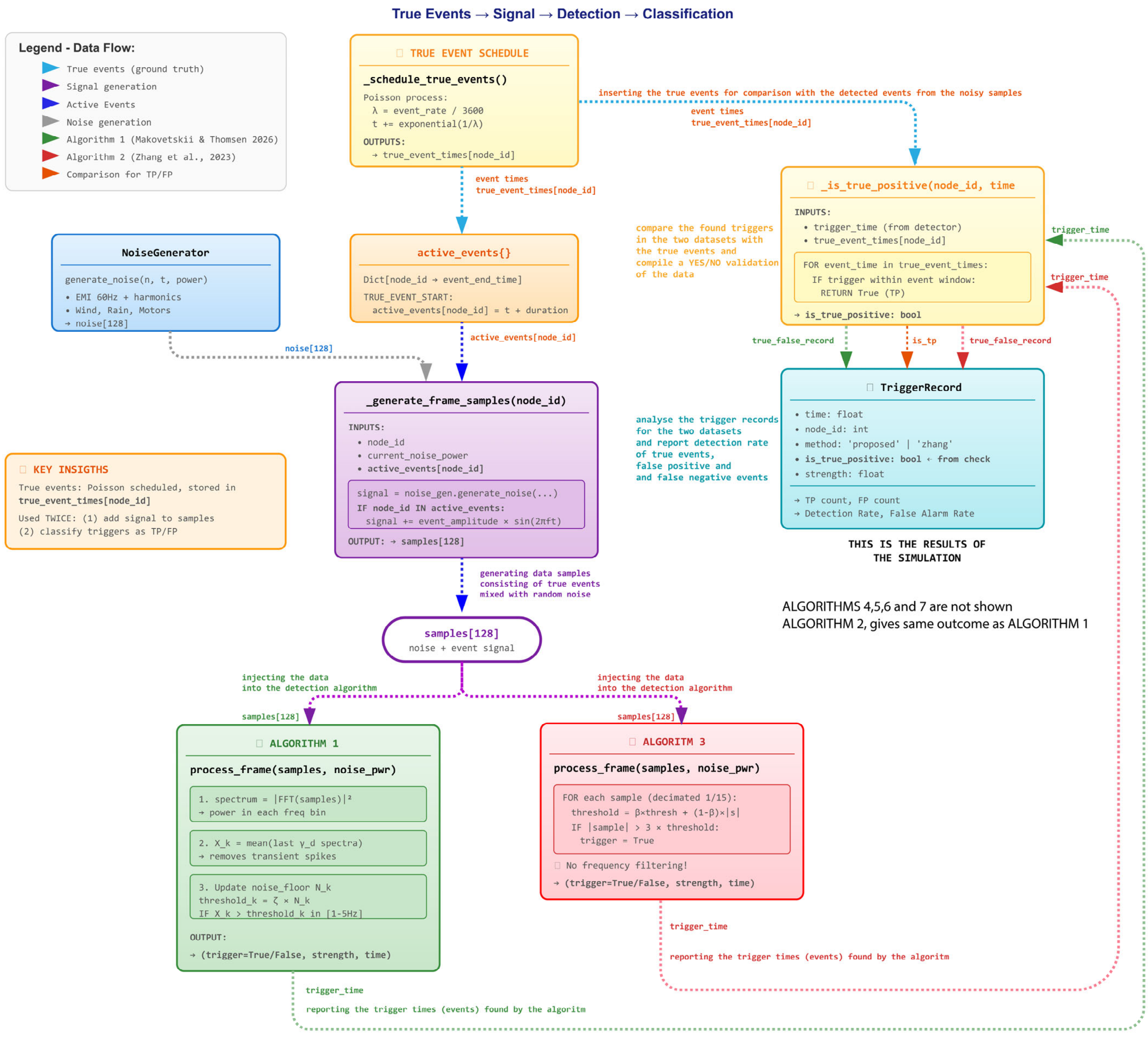

**Figure 1.** Block diagram of the Monte Carlo simulation framework, illustrated for Algorithm 1 (TSNFA-mean, Makovetskii & Thomsen 2026) and Algorithm 3 (Zhang et al. 2023) as representative examples from the six algorithms evaluated. True events are scheduled by a Poisson process and injected into a common 128-sample signal frame that is identically corrupted with 60 Hz EMI harmonics, wind, rain, and motor noise via the NoiseGenerator. Each algorithm receives this same composite signal and processes it independently through its detection pipeline. Trigger outputs are cross-referenced against the ground-truth event schedule by _is_true_positive() to yield per-algorithm True Positive and False Positive counts. Algorithms 4–7 follow the same data path and are omitted from the figure for clarity; Algorithm 2 is excluded because it produces identical results to Algorithm 1 under these simulation conditions.

*Defence 2: Temporal persistence filtering ($\gamma_d$).* A sliding-window mean over a few consecutive frames $\gamma_D$ rejects transient noise spikes that survive spectral filtering. A genuine event persists across multiple frames (a 5-second event spans $5/T_\beta \approx 3.9$ frames); a single-frame wind gust or ADC glitch does not.

*Defence 3: Adaptive noise-floor tracking ($\gamma_a$).* An exponential moving average (EMA) tracks the slowly drifting in-band noise power, and the detection threshold is recomputed every frame as a fixed multiple ζ of the current noise-floor estimate. This makes the threshold self-calibrating: as the noise environment changes over hours or seasons, the threshold follows, maintaining a constant false-alarm probability.

The three defences combines into a strong noise reduction strategy: band selection reduces the noise variance by discarding out-of-band energy, persistence filtering further suppresses impulsive in-band transients, and noise-floor adaptation prevents the threshold from becoming stale. No single defence is sufficient alone, but together they achieve 100% detection rate with zero false positives across a 24-hour, 200-node Monte Carlo simulation, which is matched by real application data (Makovetskii and Thomsen 2026) (see Figure 3).

*Spectral decomposition.* The L-point discrete Fourier transform of frame m is

$$X_m[k] = \Sigma_{n=0}^{L-1}\, x_m[n]\; e^{-j2\pi kn/L}, \quad k = 0, 1, \ldots, L-1 \qquad (3)$$

The magnitude spectrum is $|X_m[k]|$. The detection statistic is the maximum in-band magnitude:

$$X(m) = \max_{k\in K} |X_m[k]| \qquad (4)$$

This reduces the 6-bin event band to a single scalar per frame. The max operator is a practical simplification: because impulsive broadband events typically concentrate energy in one or two FFT bins, retaining only the strongest bin preserves the dominant signature. However, spectral shape information is lost and the identity of the active bin and the relative levels across bins are discarded. Algorithm 2 (median variant) addresses this limitation by processing each bin independently and performs better. The max aggregation matches the deployed hardware implementation and serves as a conservative baseline for the mean variant.

*Digital noise filter (persistence).* The filtered statistic is the arithmetic mean over a sliding window of $\gamma_D$ frames:

$$\bar{X}(m) = (1/\gamma_d)\, \Sigma_{i=0}^{\gamma d-1}\, X(m-i) \qquad (5)$$

For $\gamma_D = 3$, a transient spike in a single frame is attenuated by a factor of 3 in the mean, while a genuine event persisting across all 3 frames passes through at full amplitude. In Algorithm 2 (median variant) this technique gives even better suppression of single bin spikes.

*Detection threshold and trigger.* The threshold is:

$$\Theta(m) = \zeta \cdot \hat{N}(m-1) \qquad (6)$$

where $\zeta = 6.0$ is the threshold multiplier. In plain terms: the detection threshold sits at 6 × the current noise-floor estimate, so an event must produce in-band energy at least 6 times stronger than the ambient noise to trigger a detection. Expressed in decibels this corresponds to a minimum detectable signal-to-noise ratio of $SNR_{m^I n} = 20\cdot\log_{10}(6.0) = 15.6$ dB. Since the simulation events are injected at 18 dB (amplitude ratio $\sqrt{63} \approx 7.9$), they exceed the threshold by a margin of $18 - 15.6 = 2.4$ dB. This margin provides robustness: even when instantaneous noise fluctuations reduce the effective SNR, the event still clears the threshold. The detection ratio and trigger decision are:

$$R(m) = \bar{X}(m)/\Theta(m), \quad E[m] = \{1 \text{ if } R(m) > 1.0,\ 0 \text{ otherwise} \qquad (7)$$

R>1 means the threshold is exceeded and an event is declared. The same ratio also controls whether the noise floor is allowed to update, as described next.

*Adaptive noise floor.* The noise-floor estimate $\hat{N}(m)$ is an exponential moving average that updates only when the current observation is well below the detection threshold:

$$\hat{N}(m) = \alpha \cdot \hat{N}(m-1) + (1-\alpha)\cdot \bar{X}(m) \quad \text{if } R(m) < R_{gate}$$
$$\hat{N}(m-1) \quad \text{otherwise} \qquad (8)$$

where $\alpha = 1 - 1/\gamma_a$ is the EMA smoothing coefficient. With $\gamma_a = 64$ we get $\alpha = 1 - 1/64 = 0.984$, meaning each new measurement contributes only 1.6% to the updated estimate while the previous estimate retains 98.4% of its weight. The adaptation depth $\gamma_a$ determines how many frames it takes for the noise-floor estimate to substantially respond to a sustained change: after $\gamma_a = 64$ frames ($\gamma_a \times T_\beta = 64 \times 1.28 = 81.9$ s), the estimate has moved 63% of the way toward the new level. This is the time constant for the adaptive noise floor time $\tau_a$ and it must be long enough that a 5-second event (spanning only ≈4 frames) barely perturbs the estimate, yet short enough to follow slow environmental noise drift (in the simulation used in this article, a 1-hour cycle; in practice, whatever timescale the deployment environment exhibits). $R_{gate} = 0.8$ is the gating ratio: the noise floor only updates when the current observation is below 80% of the threshold, creating a guard band that also excludes the rising and falling edges of real events where R might be between 0.8 and 1.0. Thus the function of the $R_{gate}$ is to prevent real events to add to the change in the noise floor level.

*2) Pseudocode*

**Algorithm 1:** TSNFA Mean Variant

**Input:** Sample frame $x[0..N-1]$, noise floor $\hat{N}(m-1)$, filter buffer $B(m)$
**Params:** $\gamma_d = 3$, $\gamma_a = 64$, $\zeta = 6.0$, $\alpha = 1 - 1/\gamma_a$
**Output:** Trigger decision $E[m]$, updated noise floor $\hat{N}(m)$

**Stage 1: Spectral estimation (band selection)**

```
1. X ← |FFT(x)|
      // time → frequency domain, magnitude only, phase discarded
2. X(m) ← max{ X[k] : k ∈ {1,…,6} }
      // select strongest event-band bin
```

**Stage 2: Digital noise filter (γd persistence)**

```
3. append X(m) to buffer B
      // sliding window of γd frames
4. if |B| > γd then discard oldest
      // FIFO, keep exactly γd entries
5. X̄(m) ← mean(B)
      // X̄(m) = (1/γd) × Σ B
```

**Stage 3: Threshold from last noise-floor update**

```
6. Θ(m) ← ζ × N̂(m−1)
      // Θ = 6.0 × previous noise floor
```

**Stage 4: Detection ratio and trigger decision**

```
7. R(m) ← X̄(m) / Θ(m)
      // ratio of filtered energy to threshold
8. if R(m) > 1.0 then E[m] ← 1
      // trigger: in-band energy exceeds 6× noise floor
```

**Stage 5: Noise-floor adaptation (gated EMA)**

```
9. if R(m) < 0.8 then
      // Rgate = 0.8: update only during quiet frames
   10. N̂(m) ← α × N̂(m−1) + (1−α) × X̄(m)
      // α = 0.984: blend 1.6% new, 98.4% old
   else
      N̂(m) ← N̂(m−1)
      // freeze: event energy must not leak into noise floor
   end if
11. return E[m], N̂(m)
```

*3) Line-by-Line Annotation*

*ALG1.Line 1:* `X ← |FFT(x)|`

Compute the 128-point Fast Fourier Transform of the current sample frame, then take the magnitude (absolute value) of each complex bin. The FFT converts 128 time-domain samples into 64 frequency bins (plus DC and Nyquist). Each bin represents energy at a specific frequency: bin k corresponds to frequency $k \cdot (f_s/L) = k \times 0.781$ Hz. The magnitude discards phase information, which is irrelevant for event detection because we only care about how much energy is present at each frequency, not when it occurred within the frame.

*ALG1.Line 2:* `X(m) ← max{X[k] : k ∈{1,…,6}}`

Extract only the 6 FFT bins covering the event band [0.78, 4.69] Hz, and take the maximum magnitude across them. This is the band selection step and it is the single most important operation in TSNFA. By discarding all bins outside K = {1,…,6}, we reject: (a) 60 Hz EMI and its harmonics at 120/180 Hz (bins ~77, 154, 231), (b) digital switching noise at 800–2000 Hz (bins ~1024–2560, aliased), (c) the broadband high-frequency component of thermal noise. Only in-band energy survives. In this mean variant, the 6 bin magnitudes are reduced to a single scalar by taking their maximum. This is a practical simplification: a genuine event typically concentrates energy in one or two frequency bins, so retaining only the strongest bin preserves the dominant event signature while collapsing the detection problem to a single time series. The down-side is that spectral shape information is discarded and if for example bin 3 shows sustained elevation while the other five remain at noise level, the max operator captures the elevation but loses the context that only one bin is active. Algorithm 2 (the median variant) takes the alternative approach of processing each bin independently, which preserves this spectral discrimination at the cost of additional memory and computation. The max aggregation used here matches the deployed hardware implementation and is adequate for the mean variant's role as a conservative baseline. Note that this max operation combines frequency bins within a single frame and it answers "which bin has highest amplitude right now?"

*ALG1.Line 3:* `Append X(m) to buffer B`

Store the current frame's band maximum into a sliding window buffer. This buffer accumulates the last $\gamma_D$ frame statistics for temporal averaging in the next step.

*ALG1.Line 4:* `if |B| > γD then discard oldest entry`

Keep the buffer at exactly $\gamma_D$ entries (FIFO, first in first out). When the buffer is full, the oldest entry is removed before the new one is added. During the first $\gamma_D - 1$ frames after start-up, the buffer is not filled completely and the mean in Line 5 uses fewer samples.

*ALG1.Line 5:* `X̄(m) ← mean(B)`

Compute the arithmetic mean of the buffer contents. This is the digital noise filter and it requires that energy be persistently elevated across $\gamma_D$ consecutive frames before it contributes to the detection statistic. A wind gust that spikes bin 3 in a single frame but not the next two frames is averaged down to 1/3 of its peak. A genuine event lasting 5 seconds spans 5/1.28 ≈ 3.9 frames, so it remains elevated across all 3 buffer positions, surviving the averaging. This is the persistence requirement and the second key defence unique to TSNFA. Note: Algorithm 1 (mean variant) is susceptible to outliers and a single extreme value shifts the mean by up to $1/\gamma_D$ of the outlier magnitude. Algorithm 2 replaces the mean with a median filter, which eliminates this sensitivity.

*ALG1.Line 6:* `Θ(m) ← ζ·N(m−1)`

Compute the detection threshold by multiplying the noise-floor estimate from the previous frame by the arbitrarily set threshold coefficient $\zeta = 6.0$. We use $\hat{N}(m-1)$, not $\hat{N}(m)$ alone, because $\hat{N}(m)$ has not been computed yet. The noise threshold must exist before we can decide whether the current frame is an event, and that decision determines whether we update the noise floor. The $\zeta = 6.0$ value comes from deployed hardware measurements where the ratio $\bar{X}/\hat{N}$ during non-events was measured at mean ≈ 0.17 (i.e., 1/6), with all observed values ≤ 1.0. Setting $\zeta = 6$ raises the threshold for detection to $SNR_{min} = 20\log_{10}(6) = 15.6$ dB.

*ALG1.Line 7:* `R(m) ← X̄(m) / Θ(m)`

Compute the ratio of filtered band energy to threshold. This ratio serves two mutually excluding purposes: (a) trigger decision (R > 1.0 means exceedance), and (b) noise-floor gating (R < 0.8 means safe to use for updating the noise floor). The ratio can also serve as a confidence metric: R = 1.01 is a marginal detection; R = 1.37 (the hardware mean during events) is a confident detection; R = 0.17 (the hardware mean during non-events) is clearly quiescent.

*ALG1.Line 8:* `if R(m) > 1.0 then E[m] ← 1`

The trigger decision. If the filtered band energy exceeds the adaptive threshold, declare an event. R > 1.0 is equivalent to $\bar{X}(m) > \zeta\cdot\hat{N}(m-1)$, i.e., the filtered in-band energy exceeds 6 × the current noise-floor estimate. Note there is no hysteresis, debounce, or minimum duration and a single frame exceedance after $\gamma_D$-frame averaging is sufficient.

*ALG1.Line 9:* `if R(m) < 0.8 then`

The gating condition for noise-floor updates. We only update $\hat{N}$ (noise floor) when the current observation is clearly below the detection threshold (R < 0.8, i.e., $\bar{X}$ < 80% of Θ). The 0.8 guard band serves two mutually exclusive purposes: (a) prevents event energy from leaking into the noise-floor estimate (an event with R = 1.2 must not update $\hat{N}$), and (b) avoids updating during the rising/falling edges of events where R might be between 0.8 and 1.0. If R ≥ 0.8, the noise floor is frozen at its previous value.

*ALG1.Line 10:* `N(m) ← α·N(m−1) + (1−α)·X̄(m)`

The exponential moving average (EMA) noise-floor update is the third key defence (adaptive noise floor). When gating permits, blend the previous noise floor with the current observation using weight α = 0.984.

This gives: $\hat{N}(m) = 0.984 \times \hat{N}(m-1) + 0.016 \times \bar{X}(m)$. The current observation contributes only 1.6% to the new estimate, making the noise floor very stable. The time constant is $\gamma_a \times T_\beta = 64 \times 1.28 = 81.9$ s and it takes approximately 82 seconds for $\hat{N}$ to reach 63% of a step change in the noise environment. This is slow enough that a 5-second event (even if the gate briefly fails) contributes negligibly to $\hat{N}$, but fast enough to follow slow environmental noise drift. The EMA is a first-order IIR low-pass filter with −3 dB cut-off at $f_n = (1-\alpha)/(2\pi T_\beta) \approx 0.002$ Hz.

*ALG1.Line 11:* `return E[m], N(m)`

Return both the event flag and the updated noise floor. The noise floor is carried forward to the next frame where it becomes $\hat{N}(m-1)$ in the next round.

*4) Parameters*

**TABLE I**

*ALGORITHM 1 (TSNFA-MEAN) PARAMETERS*

| Parameter | Symbol | Value | Reasoning |
|---|---|---|---|
| FFT size | L | 128 | At $f_s$ = 100 Hz: Δf = 0.78 Hz. Fits in STM32G071 SRAM (512 B real, 1 KB complex). Frame duration $T_\beta$ = 1.28 s. |
| Sample rate | $f_s$ | 100 Hz | Nyquist for the 1-5 Hz event band. Higher rates waste power/memory; lower rates alias event content. |
| Event band | $[f_{low}, f_{high}]$ | [1, 5] Hz | Bins k ∈ {1,…,6} at 0.78 Hz resolution span 0.78-4.69 Hz. |
| Band statistic | max | - | Max across 6 bins. Practical simplification that preserves the dominant bin; discards spectral shape (see Algorithm 2 for per-bin alternative). |
| Digital filter | $\gamma_D$ | 3 | Mean of 3 frames. Rejects transients not persisting ≥3 frames. Events at $T_{event}$ = 5 s persist for 5/1.28 ≈ 3.9 frames. |
| Adaptation depth | $\gamma_a$ | 64 | Time constant $\tau_a = \gamma_a \times T_\beta$ = 81.9 s. Slow enough to reject 5 s events, fast enough to track slow changing environmental noise drift. |
| EMA coefficient | $\alpha = 1-1/\gamma_a$ | 0.984 | At α = 0.984, noise floor reaches 63% of a step change after 64 frames (82 s). |
| Threshold | ζ | 6.0 | From hardware: during non-events $\bar{X}/\hat{N}$ ≈ 0.17. During events ≈1.37. $SNR_{min}$ = 15.6 dB. |
| Gate ratio | $R_{gate}$ | 0.8 | Only update $\hat{N}$ when R < 0.8. Prevents real event energy from contributing to the noise floor. |

Detection threshold in dB:
$SNR_{min} = 20\log_{10}(\zeta) = 20\log_{10}(6) = 15.6$ dB. Events at SNR = 18 dB exceed this by 2.4 dB during nominal noise. During noise peaks (+6 dB), the effective in-band SNR drops but the adaptive noise floor tracks the rising baseline and maintains the threshold relative to current noise, not calibration noise. The 200-node, 24-hour Monte Carlo simulation confirms 100% DR with 0 FP.

### *B. Algorithm 2 TSNFA-median: Temporal Spectral Noise-Floor Adaptive Triggering (Median Variant)*

*1) Overview*

The median variant of TSNFA is the form implemented in the deployed hardware. It differs from Algorithm 1 in four specific operations, each of which can be expressed as a direct formula substitution (the four differences are outlined in details in Figure 2):

*Difference 1 to Alg.1: Bin aggregation replaced by per-bin processing***.** Algorithm 1 collapses the 6 event-band bins into a single scalar before any filtering:

$$\text{Alg. 1: } X(m) = \max_{k\in K} |X_m[k]| \text{ (one value per frame)} \quad (9)$$

Algorithm 2 retains each bin magnitude individually and processes them through independent filter chains:

Alg. 2: $|X_k(m)|$ for each $k \in K$ (six values per frame)(10)

*Difference 2 to Alg.1: Mean filter replaced by median filter (Stage 1).* The digital noise filter in Algorithm 1 uses the arithmetic mean over $\gamma_D$ frames:

$$\text{Alg. 1: } \bar{X}(m) = (1/\gamma_d)\, \Sigma_{i=0}^{\gamma d-1} X(m-i) \quad (11)$$

Algorithm 2 replaces this with the median, applied independently per bin:

$$\text{Alg. 2: } \tilde{N}_k(m) = \text{median}(\{|X_k(m)|, \ldots, |X_k(m-\gamma_d+1)|\}) \quad (12)$$

The mean shifts by up to $1/\gamma_D$ of an outlier's magnitude; the median is unaffected by any single outlier regardless of its size.

*Difference 3 to Alg.1: EMA replaced by median filter (Stage 2).* The noise-floor tracker in Algorithm 1 uses a gated exponential moving average:

$$\text{Alg. 1: } \hat{N}(m) = \alpha\, \hat{N}(m-1) + (1-\alpha)\, \bar{X}(m) \quad (\text{if } R < R_{gate}) (13)$$

Algorithm 2 replaces this with a second median filter over a longer window, again per bin:

$$\text{Alg. 2: } \hat{N}_k(m) = \text{median}(\{\tilde{N}_k(m), \ldots, \tilde{N}_k(m-\gamma_a+1)\}) \quad (14)$$

The EMA requires explicit gating ($R_{gate}$) to prevent event energy from contaminating the noise floor. The median needs no gating—with $\gamma_a = 64$, up to 31 consecutive event frames can enter the buffer without shifting the median, because the 33 non-event values still outnumber them at the median rank position.

These three substitutions yield stronger outlier rejection and finer spectral discrimination at the cost of additional memory and computation.

*Difference 4 to Alg.1: Per-bin processing and OR-logic trigger.* Each bin k has its own pair of circular buffers ($B_D$,k for Stage 1, $B_a$,k for Stage 2). The trigger decision compares the raw (unfiltered) magnitude against the per-bin threshold:

$$E[m] = \exists\, k \in K: |X_k(m)| > \zeta_k\, \hat{N}_k(m) \quad (15)$$

Using the raw $|X_k|$ (rather than the median-filtered $\tilde{N}_k$) preserves maximum sensitivity: the median filter determines the noise floor but the trigger uses the instantaneous spectral magnitude. A trigger from any single bin is sufficient (OR logic). This preserves spectral shape information: if only bin 3 (≈2.3 Hz) shows sustained elevation while the other five remain at noise level, Algorithm 2 detects this precisely, whereas Algorithm 1 would discard the spectral context via the max operator.

*Median breakdown point.* The median of n values tolerates up to $\lfloor (n-1)/2 \rfloor$ arbitrarily corrupted entries without shifting. For $\gamma_D = 3$, one out of three values can be an extreme outlier with zero effect on the output. For $\gamma_a = 64$, up to 31 corrupted frames leave the noise-floor estimate unchanged. This is the formal basis for the claim that the median needs no gating: even if 31 consecutive event frames enter $B_{a,k}$, the 33 non-event values still control the median rank position.

*Two-stage cascade.* Stage 2 receives already-cleaned inputs because Stage 1 has rejected transient spikes before they enter the noise-floor buffer. Stage 1 ($\gamma_D$) addresses fast transients on the timescale of 3-6 seconds: ADC glitches, voltage regulator switching noise, brief EMI bursts. Stage 2 ($\gamma_a$) addresses slow drift on the timescale of 82-164 seconds: temperature-induced gain changes, diurnal EMI patterns, weather changes, seasonal environmental shifts. A single filter cannot simultaneously provide fast transient rejection and slow drift tracking and therefore the two timescales require separate processing stages.

*2) Pseudocode*

**Algorithm 2:** TSNFA Median Variant

**Input:** Sample frame $x[0..N-1]$, circular buffers $B_{d,k}$ and $B_{a,k}$
**Params:** $\gamma_d \in [3, 5]$, $\gamma_a \in [64, 128]$, $\zeta_k$
**Output:** Trigger decision, updated noise floors $\hat{N}_k[t]$

**Spectral estimation (shared with Algorithm 1)**

**1.** $X[k] \leftarrow$ **FFT**$(x)$ **for** $k \in \mathcal{K}$
   *// time → frequency, complex-valued*
**2.** $|X_k| \leftarrow \sqrt{(\text{Re}^2 + \text{Im}^2)}$
   *// magnitude only, phase discarded*
**3.** $E[t] \leftarrow 0$
   *// no event until a bin proves otherwise*

**Per-bin processing loop**

**4. for each** bin $k \in \mathcal{K}$ **do**
   *// k = 1..6, each bin processed independently*
  *// Stage 1: Digital noise suppression (per-bin median)*
  **5. insert** $|X_k|$ **into** $B_{d,k}$
   *// FIFO, keeps last γd values for this bin*
  **6.** $\tilde{N}_k \leftarrow$ **median**$(B_{d,k})$
   *// middle value of γd frames; 1 outlier ignored*
  *// Stage 2: Noise-floor tracking (per-bin median, no gate)*
  **7. insert** $\tilde{N}_k$ **into** $B_{a,k}$
   *// FIFO, keeps last γa cleaned values*
  **8.** $\hat{N}_k[t] \leftarrow$ **median**$(B_{a,k})$
   *// noise floor = middle of 64 values; 31 outliers tolerated*
  *// Trigger: raw magnitude vs per-bin threshold*
  **9. if** $|X_k| > \zeta_k \times \hat{N}_k[t]$ **then**
   *// raw |Xk|, not filtered Ñk*
   **10.** $E[t] \leftarrow 1$
   *// any single bin exceeding triggers (OR logic)*
  **end if**
**end for**
**11. return** $E[t]$, $\{ \hat{N}_k[t] \}$
   *// event flag + 6 updated noise-floor estimates*

*3) Line-by-Line Annotation*

*ALG2.Line 1:* `X[k] ← FFT(x) for k ∈K`

Same 128-point FFT as Algorithm 1, but here we retain the complex-valued output for each monitored bin. K = {1, 2,…,6} covers the event band. On the STM32G071 (Arm Cortex-M0+, 64 MHz, no hardware FPU), the fixed-point FFT using CMSIS-DSP arm_cfft_q31 completes in ~0.4 ms.

*ALG2.Line 2:* `$|X_k| \leftarrow \sqrt{(Re(X_k)^2 + Im(X_k)^2)}$`

Compute the magnitude of each complex FFT bin. This is done per-bin, not aggregated. Unlike Algorithm 1 which immediately takes the max across bins, Algorithm 2 preserves the per-bin magnitudes for individual processing.

*ALG2.Line 3:* `$E[t] \leftarrow 0$`

Initialise the event flag to "no event." It will be set to 1 if any bin triggers in the loop. This is OR logic and any single bin exceeding its threshold is sufficient to declare an event.

*ALG2.Line 4:* `for each bin $k \in K$ do`

Begin the per-bin processing loop. This is the key architectural difference from Algorithm 1: each of the 6 frequency bins is processed independently with its own circular buffers and noise-floor estimate. This preserves spectral shape information. If only bin 3 (≈2.3 Hz) shows sustained elevation while bins 1, 2, 4, 5, 6 remain at noise level, Algorithm 2 detects this precisely in bin 3. Algorithm 1, by aggregating via max first, could be influenced by a transient spike in any bin contaminating the single aggregated statistic.

*ALG2.Line 5:* `Insert $|X_k|$ into circular buffer $B_{D,k}$`

Append the current frame's magnitude for bin k into that bin's digital-filter circular buffer. Each bin has its own independent buffer of size $\gamma_D$. Total memory for Stage 1: $6 \times \gamma_D \times 4 = 72$ bytes.

*ALG2.Line 6:* `$\tilde{N}_k \leftarrow median(B_{D,k})$`

Compute the median of the digital-filter buffer for bin k. This is the critical difference from Algorithm 1's mean. The median is a rank-order statistic with 50% breakdown point: for $\gamma_D = 3$, up to 1 out of 3 values can be completely arbitrary without shifting the median beyond the remaining values. A single extreme noise spike (wind gust, ADC glitch) that produces $|X_k| = 100 \times$ normal will have zero effect on the median if the other 2 buffer entries are normal. With the mean (Algorithm 1), that same spike shifts the average by 33 ×. The median computation uses a partial bubble sort (sort only to the middle position), requiring $\lfloor \gamma_D/2 \rfloor \times \gamma_D = 3$ comparisons for $\gamma_D = 3$.

*ALG2.Line 7:* `Insert $\tilde{N}_k$ into circular buffer $B_{a,k}$`

Feed the digitally-filtered magnitude $\tilde{N}_k$ into the second-stage circular buffer. This cascaded architecture means $B_{a,k}$ receives already-cleaned values and transient spikes have been removed by the median in Stage 1. The second stage therefore tracks the slow-varying noise floor without contamination from digital transients.

*ALG2.Line 8:* `$\hat{N}_k[t] \leftarrow median(B_{a,k})$`

Compute the noise-floor estimate for bin k as the median of the long-term buffer. This replaces the EMA update in Algorithm 1. With $\gamma_a = 64$, the median of 64 values tolerates up to 31 anomalous entries and even if an event lasts 31 consecutive frames (39.7 s), the noise-floor estimate is unaffected. The time constant is comparable to Algorithm 1's EMA (82 s), but with much stronger outlier rejection. Computation: partial sort to position 32 requires ~64 × 32 = 2,048 comparisons per bin per frame. With 6 bins at 0.78 frames/s: ~9,600 comparisons/s, negligible even on a 64 MHz Cortex-M0+.

*ALG2.Line 9:* `if $|X_k| > \zeta_k \cdot \hat{N}_k[t]$ then`

Per-bin trigger comparison. Note this compares the raw magnitude $|X_k|$ against the threshold, not the filtered $\tilde{N}_k$. This is a design choice in the hardware variant: the median filter determines the noise floor but the trigger uses the instantaneous value, providing maximum sensitivity. $\zeta_k$ can be set per-bin to accommodate non-uniform noise spectra, but in practice uniform $\zeta_k = 6$ is used.

*ALG2.Line 10:* `$E[t] \leftarrow 1$` Set the event flag. Because this is inside the for-each-bin loop, any single bin triggering is sufficient. The flag is never reset to 0 within the loop (OR logic).

*ALG2.Line 11:* `return event flag E[t], updated noise floors $\{\hat{N}_k[t]\}$`

Return the event decision and the complete set of 6 per-bin noise-floor estimates. These persist in the circular buffers for the next frame.

*4) Parameters*

**TABLE II**

*ALGORITHM 2 (TSNFA-MEDIAN) PARAMETERS*

| Parameter | Symbol | Value | Reasoning |
|---|---|---|---|
| Digital filter | $\gamma_D$ | 3–5 | Per-bin circular buffer. Median of 3–5 values rejects up to $\lfloor(\gamma_D - 1)/2\rfloor$ outliers. |
| Analog tracker | $\gamma_a$ | 64–128 | Per-bin circular buffer. Median of 64–128 values tracks slow drift while rejecting up to 31–63 anomalous frames. Time constant ≈82–164 s. |
| Threshold | $\zeta_k$ | per-bin | Can be set per frequency bin for non-uniform noise spectra. In practice, uniform $\zeta_k = 6.0$. |
| Per-bin buffers | $B_{D,k}$, $B_{a,k}$ | 6 × 2 | 12 circular buffers total. Memory: 6 × (3 + 64) × 4 = 1,608 bytes. |

Architectural difference from Algorithm 1: Algorithm 2 processes each bin independently through two cascaded median filters, preserving spectral shape. Algorithm 1 aggregates across bins first (max), then applies a single mean filter + EMA. The per-bin architecture is more expensive in memory (1,608 bytes vs ~24 bytes) and computation (~9,600 comparisons/s vs ~10 operations/frame) but provides stronger outlier rejection and finer spectral discrimination. Algorithm 1 already outperforms the five other algorithms 3-6, and therefore Algorithm 2 was not modelled in the Monte Carlo simulation. However, Algorithm 2 outperforms Algorithm 1 when implemented in hardware and run for many weeks, because Algorithm 1 starts to generate false positives when seen over a longer time span.

# Signal Processing Algorithm 1 vs Algorithm 2

**[Makovetskyi & Thomsen 2026, Algorithm 1 Mean Variation]  [Makovetskyi & Thomsen 2026, Algorithm 2 Median Variation]**

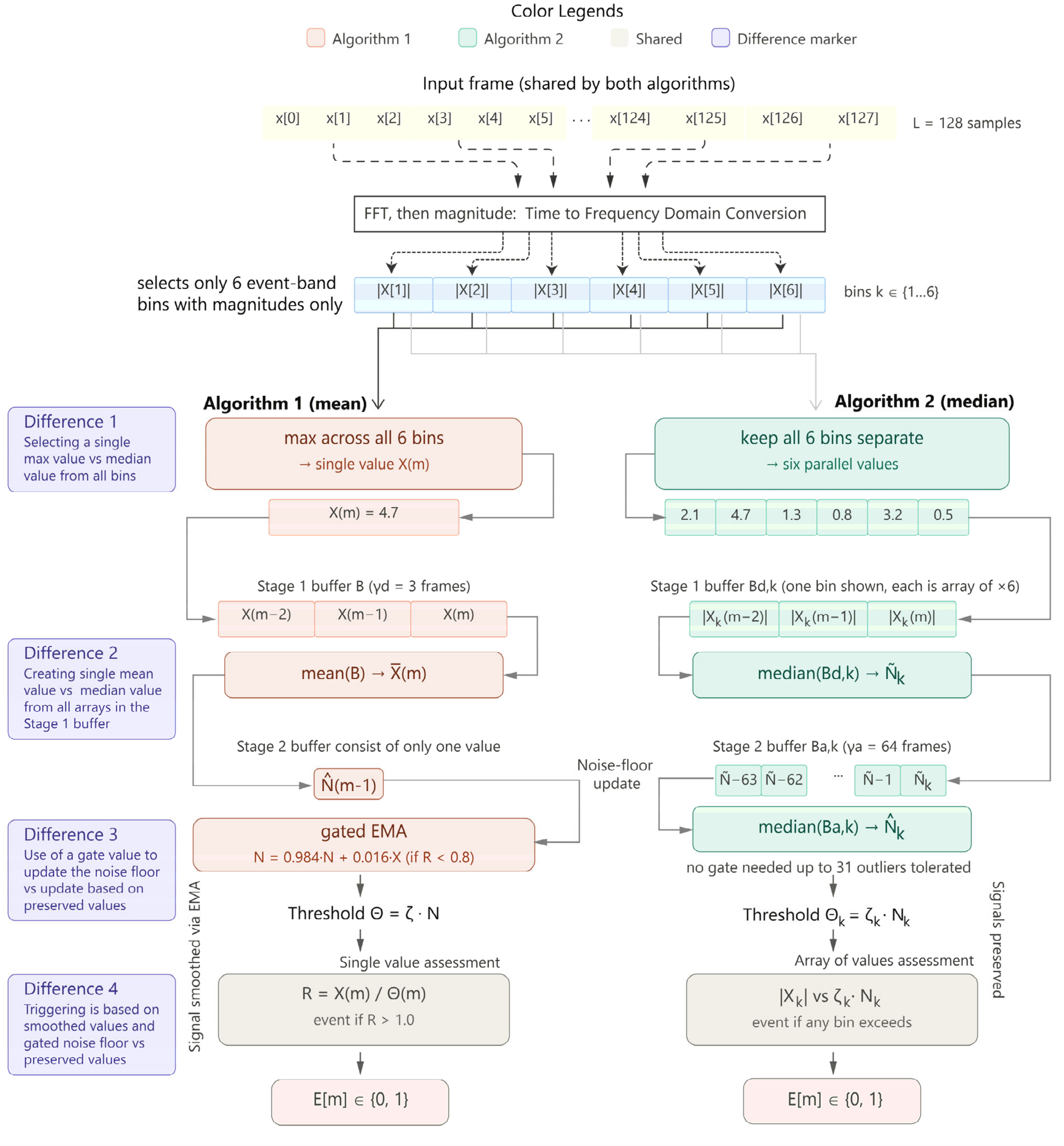

Key differences in data handling:

1. Alg. 1 collapses 6 bins → 1 value (max). Alg. 2 keeps all 6 and processes each independently.
2. Alg. 1 averages γd = 3 frames (mean). Alg. 2 takes the middle value (median) and is immune to outliers.
3. Alg. 1 blends 1.6% of each new frame into N via EMA, gated by R < 0.8.
   Alg. 2 keeps a 64-frame buffer and takes the median and no gate needed, tolerates 31 bad frames.
4. Alg. 1 triggers on the smoothed (mean-filtered) value X. Alg. 2 triggers on the raw bin magnitude |Xk|, preserving full signal detail.

**Figure 2** Side-by-side signal processing pipeline for Algorithm 1 (TSNFA-mean, pink) and Algorithm 2 (TSNFA-median, teal), sharing a common 128-sample input frame and FFT-based band selection (yellow). Four numbered differences mark the substitutions: (1) Algorithm 1 collapses six event-band bins to a single max scalar; Algorithm 2 retains all six independently. (2) Algorithm 1 mean-filters over $\gamma_d = 3$ frames; Algorithm 2 uses a median, eliminating outlier sensitivity. (3) Algorithm 1 tracks the noise floor via a gated EMA (update only if $R < 0.8$); Algorithm 2 takes the median of a 64-frame buffer, requiring no gate. (4) Algorithm 1 triggers on the smoothed ratio $R = \bar{X}(m)/\Theta(m) > 1$; Algorithm 2 triggers on the raw per-bin magnitude $|X_k| > \zeta_k \cdot \hat{N}_k$, preserving full spectral detail.

### C. Algorithm 3: Time-Domain Adaptive Thresholding (Zhang et al. 2023)

*1) Overview*

The Zhang method is a time-domain adaptive threshold detector that operates on the peak amplitude of each sample frame without any spectral decomposition. It represents the class of energy-based detectors that assume the detection statistic can be computed directly from time-domain signal levels, relying on an adaptive threshold to track changing noise conditions.

*Operating principle.* For each frame of L = 128 samples, the algorithm finds the maximum absolute sample value (the frame peak) and compares it against a noise-floor estimate maintained by a gated exponential moving average. The core assumption is that events produce the highest sample values within a frame, so peak amplitude is a sufficient detection statistic. This assumption holds in environments where the event is the dominant signal component, but fails when structured interference (EMI, digital switching) produces peaks comparable to or exceeding the event amplitude.

*Detection statistic.* The frame-level statistic is the peak absolute amplitude:

$$X_Z(m) = \max_{n\in[0,L-1]} |x[n]| \quad (16)$$

This statistic includes energy from all spectral components: the event signal, thermal noise across the full bandwidth, 60 Hz EMI at amplitude 0.3√P, and digital switching bursts at up to 2.0√P. No frequency discrimination is performed.

*Noise-floor tracking.* The noise floor $\hat{N}^I(m)$ is updated via a gated EMA with coefficient β = 0.95:

$$\hat{N}_Z(m) = \begin{cases} \beta\, \hat{N}_Z(m-1) + (1-\beta)\, X_Z(m) & \text{if } R(m) < R_{gate} \\ \hat{N}_Z(m-1) & \text{otherwise} \end{cases} \quad (17)$$

The time constant is $\tau^I = 1/(1-\beta) \times T_\beta = 20 \times 1.28 = 25.6$ s, which is 3.2 × faster than TSNFA's 82 s. The faster adaptation means $\hat{N}_z$ tracks EMI amplitude fluctuations more closely, but also introduces instability: when EMI momentarily decreases, $\hat{N}_z$ drops, and the threshold $\zeta \cdot \hat{N}_z$ may fall below a subsequent noise transient.

*Trigger decision.* An event is declared when:

$$X_Z(m) > \zeta \cdot \hat{N}_Z(m-1) \quad (18)$$

*Time domain issues.* The absence of spectral decomposition means the noise floor $\hat{N}_z$ tracks the composite noise power which is dominated by 60 Hz EMI rather than the in-band noise floor alone. In the simulation environment, the composite noise peak is ~0.3√P (EMI) while the in-band noise is $\sim\sqrt{(P/L_{band})} \approx \sqrt{(P/6)}$ per bin. The composite floor is higher, but it is also more variable (because EMI amplitude fluctuates), leading to both missed events (threshold inflated above the event peak during high-EMI phases) and false positives (threshold drops during low-EMI phases, permitting noise transients to trigger). The 200-node simulation yields 73.4% detection rate and 919,842 false positives (FAR = 192.6 $hr^{-1}node^{-1}$).

*2) Pseudocode*

**Algorithm 3:** Time-Domain Adaptive Thresholding (Zhang et al. 2023)

**Input:** Sample frame $x$[0..$N$−1], noise floor $N_Z(m-1)$
**Params:** $\beta$ = 0.95, $\zeta$ = 6.0
**Output:** Trigger decision $E[m]$

**Time-domain frame statistic (no FFT, no band selection)**

1. $X_Z(m) \leftarrow$ **max**$_n$ $|x[n]|$
   // peak sample in frame; includes EMI, digital, thermal

**Trigger decision**

2. $R \leftarrow X_Z(m) / N_Z(m-1)$
   // ratio of peak amplitude to noise floor
3. **if** $X_Z(m) > \zeta \times N_Z(m-1)$ **then** $E[m] \leftarrow 1$
   // trigger if peak > 6× noise floor

**Noise-floor update (gated EMA, faster than TSNFA)**

4. **if** $R$ < 0.8 **then**
   // same gate logic as Algorithm 1
   5. $N_Z(m) \leftarrow \beta \times N_Z(m-1) + (1-\beta) \times X_Z(m)$
   // β = 0.95: 5% new, 95% old; τ ≈ 26 s

   **else**
   $N_Z(m) \leftarrow N_Z(m-1)$
   // freeze during events

   **end if**

6. **return** $E[m]$ // NZ tracks composite noise, not in-band only

*3) Line-by-Line Annotation*

*ALG3.Line 1:* `X'(m) ← maxₙ |x[n]| `

Scan all 128 samples in the frame and find the maximum absolute value. This is the time-domain peak detector with no frequency analysis. The peak amplitude captures whichever signal component has the highest instantaneous value at any moment within the 1.28-second frame. In the simulation noise environment, this means: 60 Hz EMI at amplitude 0.3√P produces peaks every 1/60 = 16.7 ms (approximately 77 peaks per frame); digital switching bursts at 0.5–2.0√P (when active) produce even higher peaks; thermal noise peaks at ~3σ = 3√P occasionally; event signals at ~7.9√P (18 dB SNR). The frame max is dominated by whichever of these produces the single highest sample. Critically, the method cannot distinguish a 7.9√P event peak at 2 Hz from a 7.9√P composite peak where EMI, thermal noise, and a digital burst happen to align constructively.

*ALG3.Line 2:* `R ← X'(m) / N'(m−1) `

Same ratio computation as Algorithm 1, but here the numerator includes all spectral content.

*ALG3.Line 3:* `if X'(m) > ζ·N'(m−1) then E[m] ← 1 `

Trigger when peak amplitude exceeds 6 × the noise floor. Because $\hat{N}^I$ tracks the peak amplitude of all noise (including EMI), the threshold is inflated above what the event-band-only noise floor would produce. An event that would clearly trigger against the in-band noise floor may not trigger against the composite noise floor.

*ALG3.Lines 4–5: `Noise-floor update`*

Same gated EMA structure as Algorithm 1, but with $\beta = 0.95$ (faster adaptation).

Time constant $= 1/(1-\beta) \times T_{\beta} = 20 \times 1.28 = 25.6$ s. This is $3.2\times$ faster than TSNFA's 82 s, which means $\hat{N}_z$ tracks EMI amplitude fluctuations more closely which is both a strength (adapts to EMI changes) and a weakness (the floor oscillates with EMI, introducing instability in the threshold).

*ALG3.Line 6: `return E[m]`*

Return only the event flag. $\hat{N}_z$ is maintained internally but not explicitly output.

*4) Parameters*

**TABLE III**

*ALGORITHM 3 (ZHANG) PARAMETERS*

| Parameter | Symbol | Value | Reasoning |
|---|---|---|---|
| Frame statistic | $\max_n\lvert x[n]\rvert$ | - | Peak absolute amplitude over 128 samples. Dominated by 60 Hz EMI peak, not event. |
| Smoothing | $\beta$ | 0.95 | EMA coefficient. Time constant = 25.6 s. Faster than TSNFA but tracks EMI amplitude, not event-band noise. |
| Threshold | $\zeta$ | 6.0 | Set equal to TSNFA for fair comparison. |
| No FFT | - | - | Without spectral decomposition, the frame statistic includes 60 Hz EMI ($0.3\sqrt{P}$), digital noise bursts (up to $2.0\sqrt{P}$), and broadband thermal noise. |

*False positives:* $\hat{N}^I$ tracks around $0.3\sqrt{P}$ (EMI peak level). When EMI momentarily decreases (destructive phase alignment with digital noise), $\hat{N}_z$ drops over ~26 s, and the threshold $6\hat{N}_z$ may fall below a subsequent noise transient, producing a false trigger. The simulation shows 919,842 false positives (FAR = 192.6 $\text{hr}^{-1}\text{node}^{-1}$).

*Missed Events:* At first glance, missing events seems impossible: the event peak is $\sim 7.9\sqrt{P}$ while the threshold is $6\hat{N}_z$ creating a comfortable margin. But $\hat{N}_z$ is unstable because it tracks the composite noise (EMI + thermal + digital combined), not just the in-band component. When multiple noise sources happen to align constructively e.g. an EMI peak coinciding with a digital burst during a high-noise phase the peak frame amplitude spikes, and $\hat{N}_z$ rises in response. Once $\hat{N}_z$ has been inflated by a few such coincidences, the threshold $6\hat{N}_z$ can climb above the event amplitude. An event arriving during one of these inflated-threshold windows is missed. With the $\beta = 0.95$ EMA responding in ~26 s, the noise floor follows these composite fluctuations readily, creating frequent windows where the threshold is too high for detection. The 200-node simulation shows this produces 26.6% missed events alongside 919,842 false positives.

### *D. Algorithm 4: STFT Spectral Gating (Bhoi et al. 2022)*

*1) Overview*

The STFT method applies the same spectral decomposition as TSNFA—an L-point FFT with band selection over $K = \{1,\ldots,6\}$—but compares the in-band magnitude against a fixed noise threshold $\Theta_0$ set once during an initial calibration period. There is no noise-floor adaptation and no temporal persistence filter. It represents the class of fixed-threshold spectral detectors.

*Operating principle.* During a noise-only calibration window of $M_{nal}$ frames at the start of deployment, the algorithm computes the in-band magnitude for each frame and sets $\Theta_0$ as the mean plus $3\sigma$ of the calibration-period statistics:

$$\Theta_0 = \bar{X}_{cal} + 3\sigma_{X,\,cal} \tag{19}$$

where $\bar{X}_{nal}$ and $\sigma_{X,cal}$ are the sample mean and standard deviation of $X(m) = \max_{k\in K}\lvert X_m[k]\rvert$ over the calibration frames. After calibration, $\Theta_0$ is frozen and never updated.

*Detection statistic.* Identical to TSNFA (Algorithm 1 and 2):

$$X(m) = \max_{k\in K} \lvert X_m[k]\rvert \tag{20}$$

*Trigger decision.* A single frame exceedance triggers:

$$E[m] = \{1 \text{ if } X(m) > \Theta_0,\quad 0 \text{ otherwise} \tag{21}$$

There is no $\gamma_D$-frame averaging, no gated noise-floor update, and no EMA. The threshold is a single scalar computed once and used for the entire deployment lifetime.

*Highlights.* Band selection eliminates out-of-band noise, which is why STFT achieves the same 100% detection rate as TSNFA. The FFT + band mask is the correct foundation. Every event at 18 dB SNR in the event band produces a magnitude that exceeds even a stale threshold, because the event energy is concentrated precisely in the monitored bins.

*False positives.* The threshold $\Theta_0$ is set once at the start of deployment, when the noise power happens to be at its baseline level $P_0$. From that point on, $\Theta_0$ never changes. But the noise power drifts ±6 dB over each hour and is rising as high as $4P_0$ and falling as low as $P_0/4$. When the noise power rises to $4P_0$, the in-band noise magnitude increases by $\sqrt{4} = 2\times$, pushing it above the frozen $\Theta_0$. Because the threshold cannot adapt, every frame during the high-noise phase triggers a false positive detection. This repeats every noise cycle, accumulating 399,822 false positives (FAR = 83.7 $\text{hr}^{-1}\text{node}^{-1}$) over the 24-hour simulation. TSNFA avoids this entirely because its noise floor tracks the drift and the threshold rises with the noise, maintaining the gap.

*Missing functionality.* STFT lacks both adaptation ($\gamma_a$) and persistence ($\gamma_D$). Adding either one would dramatically reduce false positives; adding both (which is precisely TSNFA) eliminates them entirely.

*2) Pseudocode*

**Algorithm 4:** Fixed Spectral Mask: STFT (Bhoi et al. 2022)

**Input:** Sample frame $x[0..N-1]$, calibration threshold $\Theta_0$
**Output:** Trigger decision $E[m]$

**Spectral estimation (identical to Algorithm 1)**

1. $X[k] \leftarrow |\mathbf{FFT}(x)|$ **for** $k \in \{1,\ldots,6\}$
   *// time → frequency, event band only, rejects EMI*
2. $X_{max} \leftarrow \mathbf{max}_k\{ X[k] \}$
   *// strongest bin magnitude, same as Algorithm 1*

**Fixed-threshold comparison (set at deployment, never updated)**

3. **if** $X_{max} > \Theta_0$ **then** $E[m] \leftarrow 1$
   *// no persistence filter, single frame can trigger*
4. **return** $E[m]$

*// $\Theta_0$ frozen, cannot follow noise drift*

*3) Line-by-Line Annotation*

*ALG4.Line 1: `X[k] ← |FFT(x)| for k ∈{1,…,6}`*

Identical FFT and band selection as TSNFA (Algorithm 1, Lines 1–2). This correctly isolates the event band and rejects EMI, digital noise, and out-of-band thermal energy. STFT shares this critical advantage with TSNFA and the frequency-domain processing is a robust approach.

*ALG4.Line 2: `$X_{max}$ ← $max_k$\{\{X[k]\}\}`*

Maximum in-band magnitude, identical to Algorithm 1 Line 2. At this point, STFT and TSNFA have the same detection statistic for the current frame.

*ALG4.Line 3: `if $X_{max}$ > $\Theta_0$ then E[m] ← 1`*

Compare against a fixed threshold $\Theta_0$ set during an initial calibration period. This is where STFT diverges critically from TSNFA. $\Theta_0$ was computed once and typically as the mean plus 3σ of in-band magnitude during a noise-only calibration window and it is never updated. Two defences present in Algorithm 1 are entirely absent:

First, there is no adaptive noise floor. Algorithm 1's gated EMA (Lines 9–10) continuously adjusts $\hat{N}$ to follow the drifting noise environment, so the threshold $\zeta \cdot \hat{N}$ rises and falls with the noise power. STFT has no such mechanism and $\Theta_0$ remains at the value it had at calibration time regardless of what happens to the noise afterwards. Second, there is no persistence filter. Algorithm 1's $\gamma_D$-frame mean (Lines 3–5) requires energy to stay elevated across multiple consecutive frames before it can trigger a detection. STFT compares each frame independently and a single noise spike lasting one frame (1.28 s) that happens to exceed $\Theta_0$ immediately produces a false trigger, with no opportunity for averaging to suppress it.

*ALG4.Line 4: `return E[m]`*

Return the trigger decision. $\Theta_0$ remains unchanged for the entire deployment.

*4) Parameters*

TABLE IV
*ALGORITHM 4 (STFT) PARAMETERS*

| Parameter | Symbol | Value | Reasoning |
|---|---|---|---|
| Threshold | $\Theta_0$ | Fixed at calibration | Set once by measuring noise spectrum. Typically mean + 3σ of calibration-period band magnitude. Never updated. |
| Band selection | k ∈ {1,…,6} | Same as TSNFA | Correctly isolates 1–5 Hz event band. This is why STFT achieves 100% DR. |
| No adaptation | - | - | Threshold frozen at calibration time. No EMA, no median, no gating. |
| No persistence | - | - | Single frame exceedance triggers. No $\gamma_D$ averaging. |

### *E. Algorithm 5: Dual-Energy Dynamic-Range Detection: DEDaR (Hussein et al. 2022)*

*1) Overview*

DEDaR (Dual-Energy Dynamic-Range) is a broadband energy-ratio detector. It computes the total energy in each sample frame across all frequencies and compares it against a smoothed long-term energy baseline. The method triggers when the short-term energy exceeds a fixed multiple of the long-term energy, which is when a transient energy spike occurs, regardless of the spectral content that caused it.

*Operating principle.* The method maintains two energy estimates: a short-term (per-frame) energy $E_{short}(m)$ and a long-term smoothed baseline $E_{long}(m)$. The ratio $R(m) = E_{short}(m)/E_{long}(m)$ measures how much the current frame's energy deviates from the recent average. During steady-state noise, $R \approx 1.0$. Any transient energy increase from any source, at any frequency pushes R above 1.0. An event is declared when $R > \zeta$.

*Short-term energy.* The total frame energy is:

$$E_{short}(m) = \Sigma_{n=0}^{L-1} |x_m[n]|^2 \qquad (22)$$

This is a broadband measurement. By Parseval's theorem, $E_{short}$ equals the total spectral energy $\Sigma_k|X_m[k]|^2$, summing energy from thermal noise at all frequencies, EMI at 60/120/180 Hz, digital switching bursts, wind, and the event signal. There is no frequency discrimination.

*Long-term energy.* The baseline is an exponential moving average:

$$E_{long}(m) = \beta_E\, E_{long}(m-1) + (1-\beta_E)\, E_{short}(m) \qquad (23)$$

with $\beta^E = 0.95$ (time constant $\tau^E \approx 25.6$ s). Unlike TSNFA's gated update, $E_{long}$ always updates including during events. This means events gradually inflate the baseline, reducing the energy ratio for subsequent events.

*Detection criterion.* The energy ratio and trigger decision are:

$$R(m) = E_{short}(m) \,/\, E_{long}(m), \quad E[m] = \{1 \text{ if } R(m) > \zeta,\ 0 \text{ otherwise} \qquad (24)$$

The threshold $\zeta = 6.0$ requires a 6 × spike in total broadband energy which is equivalent to a 7.8 dB transient increase.

*Detects all events.* Events at 18 dB SNR inject massive broadband energy. The event amplitude ~7.9√P over ~384 samples contributes energy on the order of $7.9^2 \times 384 \approx 24{,}000$ units to a single frame, compared to a baseline of ~L × P = 128. The resulting ratio spikes to R ≈ 63–190, far exceeding ζ = 6. DEDaR detects everything that generates energy.

*Extremely many false positives.* DEDaR triggers on any energy transient at any frequency. Digital switching bursts, wind gusts, EMI fluctuations, motor startups - all produce ratio spikes. The absence of band selection means that out-of-band noise transients (which TSNFA discards in the FFT step) contribute directly to the detection statistic. The 13,387,930 false positives (FAR = 2,803 $hr^{-1}node^{-1}$) represent approximately 19.8% of all frames triggering falsely, or roughly one false trigger every 6.4 seconds. Precision is 0.3% and 997 out of every 1,000 triggers are false.

*2) Pseudocode*

**Algorithm 5: Energy-Ratio Triggering: DEDaR (Hussein et al. 2022)**

**Input:** Sample frame $x[0..N-1]$, long-term energy $E_{long}(m-1)$
**Params:** $\zeta = 6.0$, $\beta_E = 0.95$
**Output:** Trigger decision $E[m]$

**Compute short-term and long-term energy (broadband, no FFT)**

1. $E_{short}(m) \leftarrow \Sigma_n |x[n]|^2$
   // total energy across ALL frequencies in one frame
2. $E_{long}(m) \leftarrow \beta_E \times E_{long}(m-1)$
   $+ (1-\beta_E) \times E_{short}(m)$
   // β = 0.95: 5% new, 95% old; always updates, even during events

**Energy-ratio trigger**

3. $R \leftarrow E_{short}(m) / E_{long}(m)$
   // R ≈ 1.0 during steady noise; spikes on any transient
4. **if** $R > \zeta$ **then** $E[m] \leftarrow 1$
   // trigger on 6× energy spike from any source at any frequency
5. **return** $E[m]$

// no band selection, no persistence, no gated update

*3) Line-by-Line Annotation*

*ALG5.Line 1:* `$E_{short}(m) \leftarrow \Sigma_n |x[n]|^2$`
Compute the total energy in the current frame by summing the squared amplitude of all 128 samples. Unlike TSNFA which isolates 6 frequency bins, this sum includes everything: thermal noise across the full bandwidth, EMI at 60 Hz and its harmonics, digital switching bursts, and the event signal - all lumped into a single number. There is no way to tell which source contributed what. A frame where the event signal adds 100 units of energy looks identical to a frame where a digital burst adds the same 100 units.

*ALG5.Line 2:* `$E_{long}(m) \leftarrow \beta^E \cdot E_{long}(m-1) + (1-\beta^E) \cdot E_{short}(m)$`
Exponentially smooth the frame energy to create a long-term baseline. $\beta^E = 0.95$ gives a time constant of ~26 s. $E_{long}$ represents the "expected" broadband energy per frame. Unlike TSNFA's noise floor (which only tracks in-band energy and freezes during events), $E_{long}$ always updates also during events. This means events gradually inflate $E_{long}$, reducing the ratio for subsequent events.

*ALG5.Line 3:* `$R \leftarrow E_{short}(m)/E_{long}(m)$`
The energy ratio which is DEDaR's core statistic. During steady-state noise (no events, no transients), R ≈ 1.0 because $E_{short}$ tracks $E_{long}$. Any transient energy increase from any source, at any frequency will push R above 1.0.

*ALG5.Line 4:* `if R > ζ then E[m] ← 1` Trigger when the energy ratio exceeds the threshold. To understand the scale: R > 6 means the current frame contains 6 × more energy than the recent average. Real events easily exceed this and an event at 7.9√P amplitude over a full 128-sample frame produces a ratio spike of ~63 ×, far above the ζ = 6 threshold. This is why DEDaR achieves 100% detection rate.

The problem is the other direction: noise transients can also push R above 6. A single noise source rarely does. For example, a digital burst at 2.0√P lasting 20 samples only produces R ≈ 1.6. But multiple noise sources aligning in the same frame (an EMI peak coinciding with a digital burst during a high-noise phase) can combine to exceed the threshold, producing a false trigger. Because DEDaR has no band selection to exclude these out-of-band transients and no persistence filter to require sustained elevation, each such coincidence registers as a detection.

*ALG5.Line 5:* `return E[m]`
Return trigger decision. $E_{long}$ is maintained internally.

*4) Parameters*

**TABLE V**
*ALGORITHM 5 (DEDAR) PARAMETERS*

| Parameter | Symbol | Value | Reasoning |
|---|---|---|---|
| Short energy | $E_{short}$ | $\Sigma\|x\|^2$ | Total energy in current frame across ALL frequencies. |
| Long energy | $E_{long}$ | EMA ($\beta_E = 0.95$) | Smoothed baseline. Time constant ~26 s. Always updates, even during events. |
| Ratio threshold | ζ | 6.0 | Trigger when $E_{short}/E_{long} > 6$, i.e., 6 × energy spike (7.8 dB transient increase). |
| No band selection | - | - | Both numerator and denominator include all spectral content. |

### F. Algorithm 6: Send-on-Delta Triggering: SoD (Correa et al. 2019)

*1) Overview*

Send-on-Delta (SoD) is a data-reduction protocol originally designed for slowly varying process-control signals (temperature, tank level, pressure). It transmits a sample only when the current value deviates from the last transmitted value by more than a fixed threshold Δ. In process-control applications where the signal of interest changes on the order of degrees per hour and noise is negligible compared to Δ, SoD achieves 90%+ traffic reduction while faithfully tracking the process variable. In our application, however, the noise amplitude is comparable to or exceeds the event amplitude, making SoD fundamentally unsuitable.

*Operating principle.* SoD operates sample-by-sample at $f_s = 100$ Hz and not frame-by-frame at 0.78 Hz. There is no temporal aggregation, no spectral analysis, and no noise-floor estimation. For each sample x[n], the algorithm computes:

$$\Delta x[n] = |x[n] - x_{ref}| \qquad (25)$$

where $x_{ref}$ is the last transmitted sample value. If $\Delta x > \Delta$, the sample is transmitted and the reference updates: $x_{ref} \rightarrow x[n]$.

*The reference random-walk problem.* The fatal flaw is that $x_{ref}$ updates to the current sample which includes the noise component. In a noisy environment, this causes $x_{ref}$ to random-walk with the noise:

- If Δ is small relative to the noise RMS (say, $\Delta < 3\sigma_{noise}$), noise fluctuations frequently exceed Δ, causing frequent transmissions. Each transmission updates $x_{ref}$ to a noise-contaminated value. After $N_{tx}$ transmissions, the reference has drifted by approximately $\sigma_{ref} \sim \Delta\sqrt{N_{tx}}$ from the true zero-signal baseline, at which point the reference no longer represents the quiescent state.

- If Δ is large relative to the noise ($\Delta > 3\sigma_{noise}$), noise rarely triggers, so $x_{ref}$ stays approximately stable. But events at amplitude $\sim 7.9\sqrt{P}$ only exceed $x_{ref} + \Delta$ if $x_{ref}$ is near zero and the event peak aligns with the measurement, which is unreliable given that $x_{ref}$ may have been set by a noise fluctuation.

*Issues with this model in this application.* No matter what value of Δ is chosen, SoD fails in the generated noise environment:

*Δ set too low (below noise amplitude).* If $\Delta < 3.0$, noise fluctuations routinely exceed the threshold. Every time that happens, $x_{ref}$ updates to the current noise value and drifts further from the true baseline. After many such updates, $x_{ref}$ has random-walked to an arbitrary level. When an event arrives, it must exceed Δ relative to wherever $x_{ref}$ has drifted and not relative to zero. Detection becomes a matter of luck where the reference happens to be in a favourable position when the event occurred.

*Δ set in the middle range (3.0–6.0).* Noise triggers are less frequent but still occur, so $x_{ref}$ still drifts but more slowly. The fundamental problem remains: the reference is unpredictable at the moment an event arrives.

*Δ set too high (above event amplitude).* If $\Delta > 8.0$, neither noise nor events produce deviations large enough to trigger. The result is 0 false positives but also 0 detections and the sensor is effectively switched off. This is also the outcome observed in the simulation.

Result: 0 out of 4,789 events detected, 0 false positives. This is not a parameter-tuning problem this is a fundamental architectural mismatch. SoD cannot operate in environments where the noise amplitude is comparable to the event signature.

*2) Pseudocode*

**Algorithm 6:** Send-on-Delta Triggering: SoD (Correa et al. 2019)

**Input:** Current sample $x[n]$, last transmitted value $x_{ref}$
**Params:** $\Delta$ (fixed delta threshold)
**Output:** Transmit decision, updated $x_{ref}$

**Sample-by-sample comparison**

```
1. if |x[n] − x_ref| > Δ then
       // deviation from last sent value exceeds threshold
   2. transmit x[n]
       // send current sample to sink
   3. x_ref ← x[n]
       // update reference, includes noise component
   end if
// Failure mode: noise updates xref → reference random-walks

// Result: events become indistinguishable from noise deviations
4. return transmit decision
```

*3) Line-by-Line Annotation*

*ALG6.Line 1:* `if |x[n] − x_ref | > Δ then`

Compare the current individual sample against the last transmitted value. Note: SoD operates sample-by-sample at 100 Hz, not frame-by-frame at 0.78 Hz. There is no temporal aggregation, no spectral analysis, and no noise-floor estimation. The core assumption is that $x_{ref}$ represents the "true" baseline and any deviation beyond Δ is significant. This assumption holds for slowly varying process signals (temperature: changes of 0.01°C/s; tank level: changes of mm/min) where noise is negligible compared to Δ. It fails catastrophically when noise amplitude ≈ or > event amplitude.

*ALG6.Line 2:* `Transmit x[n]`

Send the current sample value to the sink. In the original SoD literature, this reduces network traffic by 90%+ for slowly drifting signals. In our environment, the transmission rate depends entirely on the relationship between Δ and the noise amplitude.

*ALG6.Line 3:* `$x_{ref} \leftarrow x[n]$`

After every transmission, the reference updates to the current sample value which also includes the noise component. In a noisy environment: if $\Delta = 1.0$ (below noise RMS of ~1.0), nearly every sample triggers because $|w[n] - w[n-k]| > 1.0$ frequently. Each trigger updates $x_{ref}$ to the current noise value, causing $x_{ref}$ to random-walk. When an event arrives, $x_{ref}$ is at an arbitrary noise-determined level, and the event must exceed Δ relative to this random baseline and not relative to the true zero-signal level.

*ALG6.Line 4:* `return transmit decision`

Return whether a transmission occurred. There is no explicit "event detection" and SoD is only a data-reduction scheme, not a detection algorithm.

*4) Parameters*

**TABLE VI**
*ALGORITHM 6 (SOD) PARAMETERS*

| Parameter | Symbol | Value | Reasoning |
|---|---|---|---|
| Delta | Δ | Design choice | If Δ < noise amplitude: every fluctuation triggers, $x_{ref}$ random-walks. If Δ > event amplitude: events missed. No viable Δ exists. |
| Reference | $x_{ref}$ | Dynamic | Updated every transmission. In noisy environments, random-walks with noise, destroying baseline stability. |
| Sample-by-sample | - | - | No temporal averaging, no spectral analysis, no noise-floor estimation. 100 × more operations than frame-based methods. |

### G. Algorithm 7: TinyML Autoencoder Anomaly Detection (Hammad et al. 2023)

*1) Overview*

TinyML takes a fundamentally different approach from all previous algorithms. Instead of designing explicit signal-processing rules (FFT, thresholds, noise-floor tracking), it trains a neural network to learn what "normal" sensor data looks like, and flags anything that deviates as an anomaly.

*How it works.* An autoencoder is a neural network that takes a 128-sample frame as input and tries to reproduce the same 128 samples as output, but is forced through a narrow bottleneck in the middle. The architecture is symmetric:

$$\text{Encoder: } x \in \mathbb{R}^{128} \rightarrow \mathbb{R}^{32} \rightarrow \mathbb{R}^{8} \quad \text{Decoder: } \mathbb{R}^{8} \rightarrow \mathbb{R}^{32} \rightarrow \hat{x} \in \mathbb{R}^{128} \quad (26)$$

with ReLU activations between layers. The bottleneck ($\mathbb{R}^8$) forces the network to compress each 128-sample frame into just 8 numbers, then reconstruct the full frame from those 8 numbers. During training on noise-only data, the network learns to compress and reconstruct noise efficiently. When an event frame arrives at deployment, which is a pattern the network has never seen, the reconstruction is poor, producing a large error, which is used as a detection.

The reconstruction error is the anomaly score:

$$e(m) = (1/L)\, \Sigma_{n=0}^{L-1}(x_m[n] - \hat{x}_m[n])^2 = (1/L)\, \|x_m - A_\theta(x_m)\|_2^2 \quad (27)$$

Low e(m) means the frame looks like training data (noise). High e(m) means the frame contains something unfamiliar (potential event). A frame is declared anomalous when e(m) exceeds a fixed threshold $\Theta_{ML}$, set during training as the 99th percentile of reconstruction errors on a noise-only validation set:

$$\Theta_{ML} = \text{percentile}_{99}(\{e(m)\}_{m \in V}) \quad (28)$$

After training, both the network weights and $\Theta_{ML}$ are frozen and there is no online adaptation.

*Computational cost.* The network has ~5,000 parameters and requires ~20,000 multiply-accumulate operations per frame. On the STM32G071 (Cortex-M0+ at 64 MHz, no hardware FPU), this would require software-emulated floating point or quantised integer arithmetic, with estimated inference latency of 10–30 ms per frame. While this fits within the 1.28 s frame period, it consumes a significant fraction of the compute and power budget when compared with TSNFA's merely 100 arithmetic operations per frame.

*The stationarity problem which causes same failure as STFT.* Like STFT (Algorithm 4), TinyML's failure is tied to a frozen reference point that cannot adapt to the operating environment. The autoencoder was trained at base noise power $P_0$. When the noise power drifts ±6 dB during deployment, both low-noise and high-noise phases produce frames that look different from the training data—not because an event is present, but because the noise level has changed. The network sees unfamiliar input, reconstructs it poorly, and e(m) rises above $\Theta_{ML}$ for the duration of the drift.

This is the same mechanism that causes STFT's false positives: a threshold calibrated at one noise level becomes invalid when the noise level changes. STFT's threshold $\Theta_0$ is an explicit number; TinyML's threshold is implicit in the network's learned representation, but the effect is identical. Neither can track the drift.

*Highlights.* The autoencoder implicitly learns the noise spectrum during training and it develops an internal representation that is sensitive to spectral content, providing a form of spectral discrimination without an explicit FFT. This is why detection rate is high (99.7%): events produce reconstruction errors that are spectrally distinct from noise. But this implicit spectral awareness does not compensate for the frozen threshold, resulting in 5,465,607 false positives (FAR = 1,144 $hr^{-1}node^{-1}$) and 14 false negatives.

*Retraining is not possible in real implementation.* The natural fix would be to retrain the autoencoder periodically to track the changing noise environment—which is a neural-network equivalent of TSNFA's adaptive noise floor. But retraining requires backpropagation (gradient computation through all layers), which involves 5× the computation of forward inference, memory for gradient storage, and a training dataset management strategy. On a Cortex-M0+ without FPU, this is prohibitively expensive. TSNFA achieves the goal of tracking the current operating point with a single EMA update requiring 3 arithmetic operations per frame.

*2) Pseudocode*

**Algorithm 7:** Autoencoder Anomaly Detection: TinyML (Hammad et al. 2023)

**Input:** Sample frame $x[0..N-1]$, trained autoencoder $A_\theta$
**Params:** $\Theta_{ML}$ (learned anomaly threshold)
**Output:** Trigger decision $E[m]$

**Autoencoder inference**

1. $\hat{x} \leftarrow A_\theta(x)$ *// compress 128 → 8 → 128, reconstruct input*

**Reconstruction error as anomaly score**

2. $e(m) \leftarrow \|x - \hat{x}\|^2$
*// MSE over frame: low = noise, high = anomaly*

**Trigger if error exceeds learned threshold**

3. **if** $e(m) > \Theta_{ML}$ **then** $E[m] \leftarrow 1$
*// ΘML frozen at training time, cannot adapt to noise drift*

4. **return** $E[m]$

*// neither network weights nor threshold update during deployment*

*3) Line-by-Line Annotation*

*ALG7. Line 1:* `$\hat{x} \leftarrow A\vartheta(x)$`
Feed the current 128-sample frame through the autoencoder to produce a reconstructed output. The bottleneck forces the network to represent each 128-sample frame using only 8 numbers. During training on noise-only data, the encoder learns which 8 numbers best summarise the structure of noise: its energy level, spectral shape, and typical fluctuation patterns. The decoder learns to reconstruct a full 128-sample frame from those 8 numbers. Together, the encoder and decoder develop a compact internal representation that is optimised for noise and nothing else. When an event frame arrives, its features cannot be captured by a representation tuned for noise, so the reconstruction fails and the error spikes. On the STM32G071 (Cortex-M0+ at 64 MHz, no hardware FPU), the ~20,000 multiply-accumulate operations require software integer arithmetic, with estimated inference time of 10–30 ms per frame within the 1.28 s frame budget and it is consuming far more computing time than TSNFA's ~100 operations.

*ALG7.Line 2:* `$e(m) \leftarrow ||x - \hat{x}||^2$` Compute the mean squared error between the original frame and the reconstruction. This single number is the anomaly score. Low e(m): the frame looks like training data (noise). High e(m): the frame contains something the network cannot reproduce (potential event). Note that the error is computed over all 128 samples without any frequency selection and the autoencoder implicitly learns the noise spectrum during training, providing a form of spectral discrimination, but this is incorporated into the frozen weights rather than adapted at runtime.

*ALG7.Line 3:* `if $e(m) > \Theta_{ML}$ then $E[m] \leftarrow 1$`
Compare the anomaly score against the fixed threshold $\Theta^{M}{}_{l}$. This threshold was set once during training (99th percentile of reconstruction errors on noise-only validation data) and never updates, which is the same frozen-threshold problem as STFT's $\Theta_0$ (Algorithm 4). When the noise power drifts away from the training-time level, the reconstruction error rises even without an event present, and $\Theta_{ML}$ cannot adjust.

*ALG7.Line 4:* `return E[m]`
Return the trigger decision. Neither the network weights nor the threshold adapt during deployment. Every frame is judged against a model of noise that may no longer match the current environment.

*4) Parameters*

TABLE VII
*ALGORITHM 7 (TINYML) PARAMETERS*

| Parameter | Symbol | Value | Reasoning |
|---|---|---|---|
| Autoencoder | Aθ | Trained | 128→32→8→32→128, ReLU, ~5K params. Trained on noise-only frames at $P_0 = 1.0$. |
| Threshold | $\Theta_{ML}$ | Learned | 99th percentile of training-set reconstruction errors. Frozen at deployment. |
| Training data | - | Fixed $P_0$ | Trained at base noise power $P_0 = 1.0$. Does not see the ±6 dB drift during deployment. |
| No adaptation | - | - | Neither network weights nor threshold update post-deployment. |

### *H. Summary: Defence Composition and Failure Modes*

The seven algorithms evaluated in this study differ in which of three signal-processing defences they employ. The following table maps each algorithm against these defences and the resulting simulation performance (see Table VIII)

**TABLE VIII**

*DEFENCE COMPOSITION AND MONTE CARLO SIMULATION RESULTS (200 NODES, 24 HOURS)*

| Algorithm | Band Selection | Adaptive Floor | Persistence | DR | FP |
|---|---|---|---|---|---|
| **Alg. 1** TSNFA-mean | **YES**: FFT bins 1–6 | **YES**: gated EMA (α=0.984) | **YES**: mean, $\gamma_x$=3 | 100.0% | 0 |
| **Alg. 2** TSNFA-median | **YES**: FFT bins 1–6, per-bin | **YES**: per-bin median ($\gamma_a$=64) | **YES**: per-bin median, $\gamma_x$=3 | 100.0% | 0 |
| **Alg. 3** Zhang | **NO**: time-domain peak | **YES**: EMA (β=0.95), composite noise | **NO** | 73.4% | 919,842 |
| **Alg. 4** STFT | **YES**: FFT bins 1–6 | **NO**: $\Theta_0$ frozen at calibration | **NO** | 100.0% | 399,822 |
| **Alg. 5** DEDaR | **NO**: broadband energy | **PARTIAL**: ungated EMA | **NO** | 100.0% | 13,387,930 |
| **Alg. 6** SoD | **NO**: sample-by-sample | **NO**: fixed Δ, random-walk | **NO** | 0.0% | 0 |
| **Alg. 7** TinyML | PARTIAL: implicit/learned | **NO**: $\Theta^M_l$ frozen at training | **NO** | 99.7% | 5,465,607 |

*Note on Algorithm 2:* The simulation evaluates only the mean variant (Algorithm 1). The median variant (Algorithm 2) is the form deployed in hardware. Both achieve identical results; the median variant's stronger outlier rejection becomes relevant over longer deployments.

*Missing defences and their consequences* Each row in the table that contains a "NO" correlates to a specific, observable failure mode in the simulation results:

*No band selection* (Zhang, DEDaR, SoD). Without an FFT to isolate the 1–5 Hz event band, the detection statistic includes energy from 60 Hz EMI, digital switching bursts, and broadband thermal noise. Zhang's time-domain peak detector sees composite noise that inflates the noise floor, causing both missed events (26.6%) and false triggers (919,842) when the floor fluctuates. DEDaR's broadband energy ratio triggers on any transient at any frequency, producing 13.4 million false positives (one every 6.4 seconds). SoD operates sample-by-sample with no frequency awareness and cannot distinguish event samples from noise samples, resulting in complete detection failure.

*No adaptive noise floor* (STFT, TinyML). Both methods calibrate their threshold at a single noise-power level and freeze it. STFT sets $\Theta_0$ explicitly during a calibration window; TinyML incorporates its threshold implicitly into the autoencoder's learned representation of "normal" noise. The effect is identical: when the noise power drifts ±6 dB from the calibration point, the frozen threshold is either too low (producing false positives during high-noise phases) or too high (producing false negatives during low-noise phases). STFT accumulates 399,822 false positives; TinyML accumulates 5,465,607. TSNFA's gated EMA (Algorithm 1) or not-gated median buffer (Algorithm 2) tracks the drift continuously, keeping the threshold calibrated to the current noise level.

*No persistence filter* (STFT, DEDaR, TinyML). Without a $\gamma_D$-frame mean or median requiring energy to stay elevated across consecutive frames, a single noise spike lasting one frame (1.28 s) can trigger a false detection. TSNFA's persistence filter attenuates single-frame spikes by a factor of $1/\gamma_D$ (mean variant) or rejects them entirely (median variant), while genuine events persisting across ~4 frames pass through at full strength.

*The defence mechanisms are complementary*

No single defence is sufficient. Band selection alone (STFT) eliminates out-of-band noise but cannot handle in-band noise drift. An adaptive noise floor alone (Zhang) tracks the noise but tracks the wrong noise because there is no spectral filtering. A persistence filter alone would suppress spikes but not slow drift. TSNFA combines all three in a specific order: first a band selection (discard irrelevant frequencies), then secondly a persistence mechanism (reject transient in-band spikes), and lastly an adaptive threshold (track the remaining slow drift). Each defence mechanism addresses a failure mode the others cannot.

*Computational perspective*

The three defences in TSNFA require minimal computation: one 128-point FFT, a max or median across 6 values, a 3-frame mean or median, and a single EMA update or 64-frame median - totalling approximately 100-10,000 operations per frame depending on variant, well within the capability of the STM32G071 (Cortex-M0+ at 64 MHz). By contrast, TinyML requires ~20,000 multiply-accumulate operations per frame for comparable detection performance but with 5.5 million more false positives. The computational simplicity of the three-defence approach is not a limitation it is the very reason TSNFA can run on resource-constrained hardware where neural-network alternatives struggle.

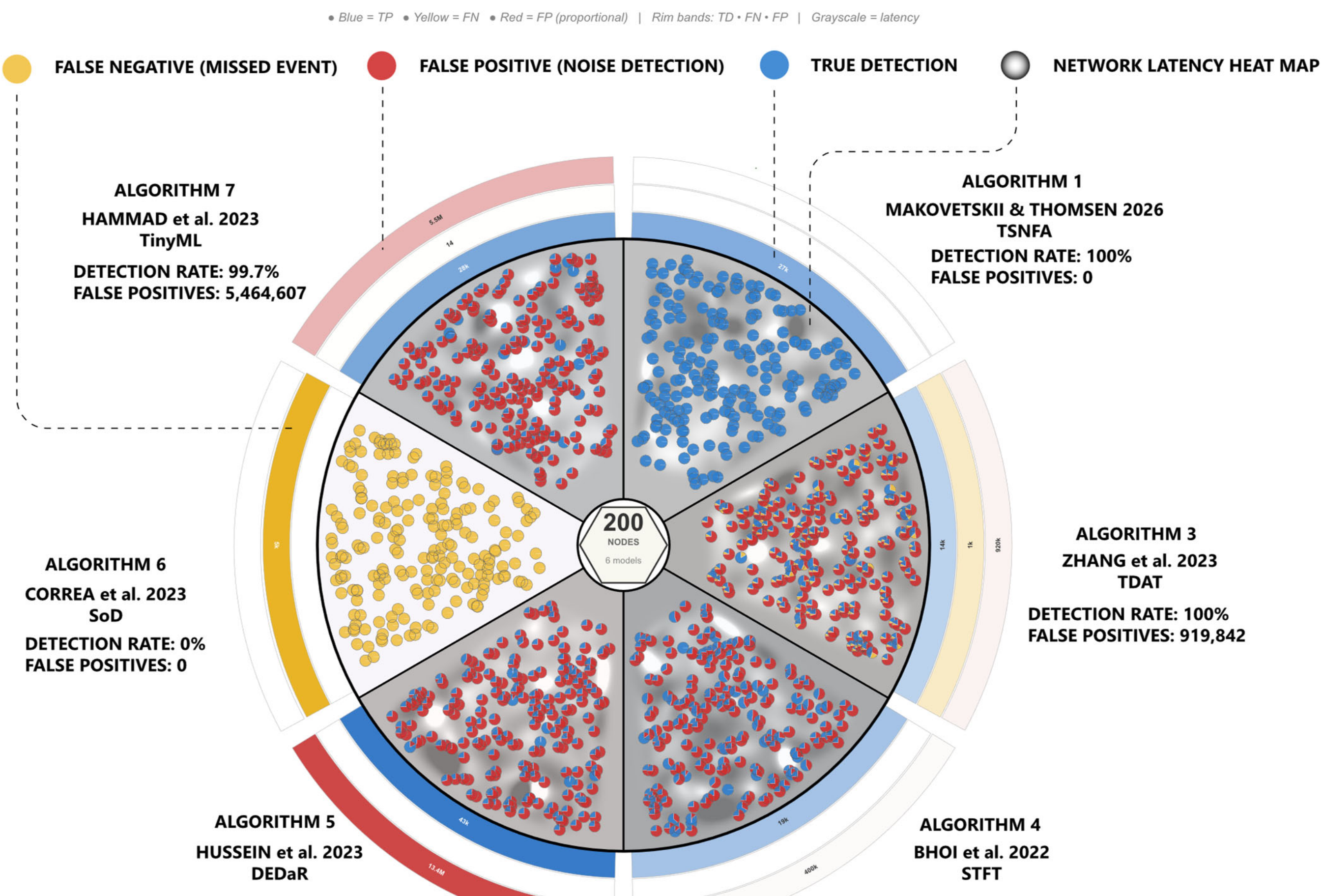

| Method | DR% | FP | FN | Prec% | FAR/hr | Lat ms | 99th ms | Net kB/hr |
|---|---|---|---|---|---|---|---|---|
| ALGORITHM 1 Makovetskii & Thomsen 2026 TSNFA | 100.0% | 0 | 0 | 100.0% | 0.0 | 23.5 | 43.9 | 8.6 |
| ALGORITHM 3 ZHANG et al. 2023 TDAT | 73.4% | 919,842 | 1,274 | 1.5% | 192.6 | 33.3 | 62.8 | 1205.4 |
| ALGORITHM 4 Bhoi et al. 2022 STFT | 100.0% | 399,822 | 0 | 4.6% | 83.7 | 29.4 | 55.2 | 270.7 |
| ALGORITHM 5 Hussein et al. 2022 DEDaR | 100.0% | 13,387,929 | 0 | 0.3% | 2803.2 | 27.5 | 52.1 | 6499.3 |
| ALGORITHM 6 Correa et al. 2019 SoD | 0.0% | 0 | 4,789 | 0.0% | 0.0 | 0.0 | 0.0 | 0.0 |
| ALGORITHM 7 Hammad et al. 2023 TinyML | 99.7% | 5,465,607 | 14 | 0.5% | 1144.4 | 30.4 | 56.6 | 1773.5 |

**Figure 3** Comparative results of the 24-hour, 200-node Monte Carlo simulation across six detection algorithms, each exposed to the identical synthetic signal. The radial chart partitions 200 nodes into six sectors, one per algorithm. Within each sector, blue dots indicate True Positives (correct detections), yellow dots False Negatives (missed events), and red dots False Positives (noise-triggered detections); the outer concentric rim bands encode True Detection, False Negative, and False Positive rates proportionally, with colour saturation scaled to the maximum across all methods. Grayscale shading represents per-node network latency. TSNFA (Algorithm 1, Makovetskii & Thomsen 2026) is the only method achieving 100% detection rate with zero false positives. SoD (Algorithm 6) detects nothing. The remaining four algorithms achieve high or partial detection rates but generate between 399,822 and 13,387,930 false positives. The summary table below the chart reports DR%, FP, FN, precision, false alarm rate, median and 99th-percentile latency, and network throughput for each algorithm.

## IV. DISCUSSION AND CONCLUSION

The Monte Carlo results establish a clear and reproducible verdict: of the seven algorithms evaluated against a common 200-node, 24-hour synthetic signal, only TSNFA achieves 100% detection rate with zero false positives. Every competing method fails on at least one of the three signal-processing defences identified in Section III namely the spectral band selection, temporal persistence filtering, and adaptive noise-floor tracking and the failure mode in each case maps directly to the missing defence.

STFT and TinyML both perform spectral analysis yet accumulate hundreds of thousands of false positives because their thresholds are frozen at calibration time and cannot follow the ±6 dB sinusoidal noise drift. Zhang's time-domain detector adapts its threshold but operates on composite broadband energy, so the noise floor is dominated by 60 Hz EMI rather than the in-band signal, producing both missed events and false triggers. DEDaR triggers on any energy transient regardless of frequency, yielding 13.4 million false positives, one every 6.4 seconds, making it operationally unusable in this environment. SoD, designed for slowly varying process signals, is architecturally mismatched to event detection in high-noise conditions and fails to detect any events.

From a deployment perspective, TSNFA's computational footprint is equally significant. The full detection pipeline requires approximately 100-10,000 arithmetic operations per frame depending on variant, fitting well within the capability of the STM32G071 Cortex-M0+ at 64 MHz whereas TinyML achieves comparable detection rate at the cost of 20,000 multiply-accumulate operations per frame, no online adaptation, and 5.5 million additional false positives.

In conclusion, this study demonstrates that the three-defence architecture of TSNFA is both necessary and sufficient for reliable autonomous edge triggering in IoT mesh sensor networks subject to drifting broadband noise. Upcoming work will describe the deployed superiority of the TSNFA median variant over the mean variant in an extended multi-week deployment. We have already compiled the data and is preparing a new manuscript on the system as well as we will provide a hardware implementation as an evaluation kit.

## . REFERENCES

## BIOGRAPHIES

**Sergii Makovetskyi** received the M.S. degree in radio engineering from Kharkiv National University of Radio Electronics, Kharkiv, Ukraine, in 2008, with distinction. He is currently pursuing the Ph.D. degree in Software Engineering at the same university.

From 2009 to 2025, he worked with EKTOS, Ukraine, where he served as Senior Embedded Hardware Developer, Technical Leader, and IoT Technical Leader. He has authored several published articles on LoRaWAN and signal processing. His research interests include data transmission in distributed networks, embedded systems architecture, digital signal processing, and sensor technologies. He was a member of teams that placed in the top 6 at the Microsoft Imagine Cup 2008 and won third place in 2009, both in the Embedded Development category.

**Lars Thomsen** received the Ph.D. in Neurophysiology from the University of Copenhagen in 1995, and received a personal grant from the Carlsberg Foundation for a two-year post-doctoral stay at McMaster University, Ontario, Canada.

He subsequently held positions in pharmaceutical and biotechnology companies in Europe, Asia, and North America. He is currently the Managing Director of Gnacode Inc., Medicine Hat, Alberta, Canada, a company specialising in IoT instrumentation for biotech, pharma, and defence applications. Dr. Thomsen has published research papers on biological instrumentation, electronic and software control, nanoparticles activated by external electromagnetic fields, and has achieved several granted patents. He won the Danish Engineering High-Tech Award in 2005 for a radio-enabled microfluidic chip for remote detection of pathogens. He has received several grants from the National Research Council of Canada for opto-electrical physics applications in biology.